\documentclass[12pt,preprint]{aastex}

\slugcomment{To appear in ApJ}

\shorttitle{Probing with Microlensing the Halos of Lens Galaxies}
\shortauthors{Mediavilla et al.}

\begin{document}



\title{Microlensing-Based Estimate of the Mass Fraction in Compact Objects in Lens Galaxies}


\author{E. MEDIAVILLA\altaffilmark{1},  J.A. MU\~NOZ\altaffilmark{2},  E. FALCO\altaffilmark{3}, V. MOTTA\altaffilmark{4}, E. GUERRAS\altaffilmark{1}, H. CANOVAS\altaffilmark{1}, C. JEAN\altaffilmark{2}, A. OSCOZ\altaffilmark{1}, A.M. MOSQUERA\altaffilmark{2}}





\altaffiltext{1}{Instituto de Astrof\'{\i}sica de Canarias, V\'{\i}a L\'actea S/N, 38200 - La Laguna, Tenerife, Spain.}
\altaffiltext{2}{Departamento de Astronom\'{\i}a y Astrof\'{\i}sica, Universidad de Valencia,  46100 - Burjassot, Valencia, Spain.}
\altaffiltext{3}{Smithonian Astrophysical Observatory, FLWO, P.O. Box, 97, Amado, AZ 85645, USA.}
\altaffiltext{4}{Departamento de Fisica y Astronomia, Facultad de Ciencias, Universidad de 
Valparaiso, Avda. Gran Breta\~na 1111, Valparaiso, Chile.}


\begin{abstract} 

We estimate the fraction of mass that is composed of compact objects in gravitational lens galaxies. This study is based on microlensing measurements (obtained from the literature) of a sample of 29 quasar image pairs seen through 20 lens galaxies. We determine the baseline for no microlensing magnification between two images from the ratios of emission line fluxes. Relative to this baseline, the ratio between the continua of the two images gives the difference in microlensing magnification. The histogram of observed microlensing events peaks close to no magnification and is concentrated below 0.6 magnitudes, although two events of high magnification, $\Delta m \sim 1.5$, are also present. We { study the likelihood of the microlensing measurements using} frequency distributions obtained from simulated microlensing magnification maps for different values of the fraction of mass in compact objects, $\alpha$.  The { concentration of microlensing measurements close to} $\Delta m \sim 0$  can be {explained} only by simulations corresponding to very low values of $\alpha$ (10\% or less). {A maximum likelihood test yields $\alpha=0.05_{-0.03}^{+0.09}$ }(90\% confidence interval) {for a quasar continuum source of intrinsic size $r_{s_0}\sim 2.6 \cdot 10^{15} \rm\, cm$}. This estimate is valid in the $0.1 - 10M_\odot$ range of microlens masses. {We study the dependence of the estimate of $\alpha$ with $r_{s_0}$, and find that $\alpha \lesssim 0.1$ for $r_{s_0}\lesssim 1.3 \cdot 10^{16} \rm\, cm$. High values of $\alpha$ are possible only for source sizes much larger than commonly expected ($r_{s_0}>>2.6 \cdot 10^{16} \rm\, cm$).} Regarding the current controversy about Milky Way/LMC and M31 microlensing studies, our work supports the hypothesis of a very low content in MACHOS (Massive Compact Halo Objects). In fact, according to our study, quasar microlensing probably arises from the normal star populations of lens galaxies and there is no statistical evidence for MACHOS in the dark halos.

\end{abstract}




\section{Introduction}
The composition of matter in the halos of galaxies is a central problem in astrophysics. During the last 10 years, several observational projects have used gravitational microlensing \citep{1986ApJ...304....1P} to probe the properties of the halos of the Milky Way (MACHO, \citealp{2000ApJ...542..281A}; EROS, \citealp{2007A&A...469..387T}) and M31 (POINT-AGAPE, \citealp{2005A&A...443..911C}; MEGA, \citealp{2006A&A...446..855D}). These experiments are based on the detection of magnification in the light-curve of a source induced by an isolated point-like (or binary) object passing near the observer's line of sight. From the successful detection of a number of microlensing events these collaborations have estimated the fraction of the halo mass that is composed of lensing objects, $\alpha$. However, the reported results disagree. For the Milky Way's halo the measurements of the MACHO collaboration \citep{2000ApJ...542..281A} correspond to a halo fraction of $0.08<\alpha<0.50$ while EROS \citep{2007A&A...469..387T} obtains $\alpha<0.08$. On the other hand, re-analysis of publicly available MACHO light-curves \citep{2004MNRAS.352..233B} leads to results similar to those reported by EROS (however, see also the counter-report by \citealp{2005MNRAS.359..464G}). For M31 the AGAPE \citep{2005A&A...443..911C} collaboration finds a halo fraction in the range $0.2<\alpha<0.9$, while MEGA \citep{2006A&A...446..855D} finds a limit of $\alpha<0.3$.

The method applied to the Milky Way and M31 { can be extended} to the extragalactic domain by observing the microlensing induced by compact objects in the lens galaxy halo in images of multiply imaged quasars { (quasar microlensing; \citealp{1979Natur.282..561C}, see also the review by \citealp{2006glsw.book..453W}). Interpreting the light-curves of QSO 2237+0305, \citet{1991AJ....102.1939W} suggest that the monitoring of microlensing variability can provide a measure of the optical depth in compact objects and in the smooth mass distribution. \citet{1996MNRAS.283..225L} proposed a statistical approach to the determination of the mass density in compact objects based on the comparison between the observed and simulated magnification probability distributions. Microlensing can also be measured from a single-epoch snapshot of  the anomalous flux ratios induced by this effect between the images of a lensed quasar (\citealp{1995ApJ...443...18W}; see also \citealp{2002ApJ...580..685S}). \citet{2004IAUS..220..103S} explore the practical application of this idea by using a sample of eleven systems with measured flux anomalies. Other quasar microlensing studies of interest for the present study are aimed at the determination of accretion disk sizes (e.g., the studies based in relatively large samples by \citealp{2007ApJ...660..146P}, \citealp{2007arXiv0707.0305M} and references therein).}

In practice, { the study of extragalactic microlensing} meets significant obstacles, in particular (e.g., \citealp{2004ApJ...605...58K}) larger time-scales for microlensing variability and lack of a baseline for no magnification needed to detect and to quantify microlensing (see, however, the time variability based studies of several individual systems in \citealp{2008ApJ...689..755M} and references therein). In addition, microlensing by an isolated object is not a valid approximation. Microlensing at high optical depth should be modeled (e.g., by simulating magnification maps; see \citealp{1992grle.book.....S}).

We avoid these obstacles by setting the baseline of no microlensing magnification using the narrow emission lines (NELs) in the spectra of lensed quasar images (\citealp{2004IAUS..220..103S} follow a similar approach but using theoretical models to define the baseline).  It is generally expected that the regions where NELs originate are very large (compared with the continuum source) and are not affected by microlensing (this assumption can also be adopted, to some extent, for low ionization broad emission lines; \citealp{2000ApJ...533..631K}; \citealp{2002ApJ...576..640A}). { If we define the baseline from emision lines measured in the same wavelength regions as the continua affected by microlensing, we can also remove the extinction and isolate the microlensing effects. 

``Intrinsic'' flux ratios between the images in the absence of microlensing can be determined from the observation of the mid-infrared and radio-emitting regions of quasars that should also be large enough to average out the effects of microlensing (see \citealp{2004ApJ...605...58K} and references therein). However, the extinction at mid-infrared and radio wavelengths is lower than the extinction at the wavelengths in which microlensing is usually detected and measured (optical, near-infrared, and X-ray). Consequently, the difference between the mid-infrared (radio) and the optical (X-ray or near-infrared) continuum fluxes will include not only the effects of microlensing but also the effects of extinction. In addition, note that the availability of data at optical wavelengths is considerably greater than at other wavelengths.}

Thus { we will use the NEL and} continuum flux ratios among the different images of a lensed QSO to estimate the difference of microlensing magnification between the images at a given epoch with certain restrictions that we detail in the following paragraphs. 

The flux (in magnitudes) of an emission line observed at wavelength $\lambda$ of image $i$ of a multiply imaged quasar is equal to the flux of the source, $m_0^{lin}\left({\lambda\over 1 +z_s}\right)$, magnified by the lens galaxy (with a $\Phi_i$ magnification factor; $\mu_i=-2.5\log \Phi_i$) and corrected by the extinction of this image caused by the lens galaxy, $A_i\left({\lambda\over 1 +z_l}\right)$ (see, e.g., \citealp{2004ApJ...605..614M}),

\begin{equation}
m_i^{lin}(\lambda)= m_0^{lin}\left({\lambda\over 1 +z_s}\right) + \mu_i + A_i\left({\lambda\over 1 +z_l}\right), 
\end{equation}

\noindent where $z_s$ and $z_l$ are the redshifts of the source and the lens, respectively.

In the case of the continuum emission, we must also take into account the intrinsic variability of the source combined with the delay in the arrival of the signal, which is different for each image, $\Delta t_i$, and the microlensing magnification, which depends on wavelength and time (with a $\phi_i\left[{\lambda\over 1 +z_l},t \right]$ magnification factor; $\Delta\mu_i=-2.5\log \phi_i$), 

\begin{equation}
m_i^{con}(\lambda,t)= m_0^{con}\left({\lambda\over 1 +z_s},t-\Delta t_i\right) + \mu_i + A_i\left({\lambda\over 1 +z_l}\right) + \Delta\mu_i \left({\lambda\over 1 +z_l},t\right). 
\end{equation}

Thus, the difference between continuum and line fluxes { cancels the terms corresponding to intrinsic magnification and extinction ($\mu_i + A_i$)}:

\begin{equation}
 m_i^{con}(\lambda,t) - m_i^{lin}(\lambda)= m_0^{con}\left({\lambda\over 1 +z_s},t-\Delta t_i\right) - m_0^{lin}\left({\lambda\over 1 +z_s}\right) + \Delta\mu_i\left({\lambda\over 1 +z_l},t\right). 
\end{equation}
 
If we consider a pair of images, $1$ and $2$, the continuum ratio relative to the zero point defined by the emission line ratio  can be written (in magnitudes) as,

\begin{eqnarray}
\Delta m (\lambda,t)= (m_1-m_2)^{con}-(m_1-m_2)^{lin}=&{\Delta\mu_1\left({\lambda\over 1 +z_l},t\right)- \Delta\mu_2\left({\lambda\over 1 +z_l},t\right) } \label{eq_baseline} \\
&+\Delta m_0^{con}\left({\lambda\over 1 +z_s},\Delta t_1-\Delta t_2\right).\nonumber 
\end{eqnarray}

We have referred the equations for the magnification of both images to an arbitrary time, $t$ (note that microlensing-induced variability between a pair of images is uncorrelated).  

The first term of Equation (\ref{eq_baseline}) is the relative microlensing magnification between images 1 and 2.  The significance of the second term, $\Delta m_0^{con}$, which represents the source variability, can be estimated by comparing the intrinsic quasar variability on time-scales typical of the time delay between images in gravitational lens systems with the expected distribution of microlensing magnifications. As we shall  discuss in Section \ref{sect_conclusions}, the intrinsic source variability is not significant for our computations.  

In summary, with the proposed method similar information as in the Milky Way MACHO experiments is obtained but with a single-epoch measurement. The objective of this study is to apply this method to published data of quasar microlensing. In \S \ref{sect_observ} we collect the data from the literature and fit models to the systems of multiply imaged quasars to derive suitable values of the projected matter density at the image locations. Using these values, probabilistic models for microlensing magnifications are derived in 
Section \ref{sect_maps} for a range of fractions of mass in compact objects. Sections \ref{sect_mle} and \ref{sect_scaling} are devoted to estimate this fraction. Finally, in \S \ref{sect_conclusions} we present and discuss the main conclusions.

\section{Observed Microlensing Magnifications and Macro-Lens Models\label{sect_observ}}

We collected the data, $\Delta m$ (see Eq. \ref{eq_baseline}), examining all the 
optical spectroscopy\footnote{ There are also several X-ray events in the literature that have been 
explained in terms of microlensing { (e.g., \citealp{2007ApJ...661...19P} and references 
therein)}. These events probably arise from a tiny inner region, as compared with the optical continuum 
emitting region, and deserve an analogous but separate study when a sufficiently large sample of 
X-ray microlensing measurements become available. }  found in the literature (see 
Table \ref{table0}). In most cases the microlensing magnification or the scaling of the emission 
line ratio with respect to the continuum ratio are directly provided by the authors { or can be 
estimated from a figure}. For SDSS 0806+2006, FBQ 0951+2635, SDSS J1001+5027, QSO 1017-207, SDSS J1206+4332, HE 1413+117, 
and SBS 1520+530 we used the electronically available or digitized spectra of the images 
to estimate the microlensing magnification following the steps described in \citet{2005ApJ...619..749M}. 
{ In Table \ref{table0} we include (when available) the flux ratios for each line and its 
corresponding continuum. Specific details of the procedure followed to obtain the data are also given.} 

{ For some of the image pairs ($\sim$30\% of the sample) there are mid-IR flux ratios available. 
Except for one system, SDSS J1004+4112 (where image C is probably affected by extinction, 
\citealp{2006ApJ...645L...5G}), they are in very good agreement with the emission-line 
flux ratios (see Table \ref{table3}). The average difference between mid-IR and emission line 
flux ratios is  only 0.11 magnitudes (0.07 magnitudes if SDSS J1004+4112 is removed). In 
fact, the agreement is unexpectedly good taking into account the possible influence of 
extinction and source variability. In any case, this comparison supports the consistency of the 
basic hypothesis (that the emission line fluxes are not affected by microlensing) and the reliability of the data.} 

Figure \ref{fig_histo_A} shows the frequency of observed microlensing magnifications, 
$f_{obs}(\Delta m)$.  This histogram exhibits two significant characteristics: 
the relatively high number of events with low or no microlensing magnification and 
the concentration ($\sim$80\%) of the microlensing events below $|\Delta m|=0.6$. Any 
model attempting to describe microlensing magnification should account for these
features. At a lower level of significance, the presence of two events of high 
magnification, $\Delta m \sim 1.5$, should also be noted. The data presented in Table \ref{table0} 
come from many different bibliographic sources with the subsequent lack of information about 
measurement procedures and estimate of uncertainties. However, even with this limitation, the low 
frequency of high magnification microlensing events in the optical seems to be a reliable observational result.

For each of the gravitational lens systems in Table \ref{table1} we have used a 
singular isothermal sphere plus external shear model (SIS+$\gamma_e$) to estimate 
values of the total projected matter density $\kappa$ and the shear $\gamma$  at the 
locations of the images (see Table \ref{table1}). The models have been computed with the ``lensmodel'' code by 
\citet{2001astro.ph..2340K} to fit the positions of the images
(CASTLES, http://www.cfa.harvard.edu/castles/). { For double systems we have used the emission-line flux ratios between images as an additional constraint.}

In Table \ref{table2} we show $n_{\kappa_1,\kappa_2}$ ($\kappa_1<\kappa_2$), 
the frequency distribution of image pairs that occur at combined projected matter densities $(\kappa_1,\kappa_2)$.
The distribution peaks at bin $(\kappa_1=0.45,\kappa_2=0.55)$.
In many of the image pairs in Table \ref{table1} the images are roughly located at similar 
distances from the lens galaxy center, $r_1\sim r_2$. In an SIS model the convergence for each of the lensed images is given by $\kappa_1=1/2(1+x)$ and
$\kappa_2=1/2(1-x)$, where $x$ is the position of the source  in units of the
Einstein radius. The image configuration $r_1\sim r_2$ is obtained when $x\gtrsim0$;  
therefore, the expected values for the convergence are $\kappa_1\lesssim0.5$ and 
$\kappa_2\gtrsim0.5$. This is in agreement with Table \ref{table2} and
in fact this simple reasoning could have been used to estimate, from a statistical
point of view, the peak of the distribution of convergence values, $n_{\kappa_1,\kappa_2}$, in our sample. 

{Observational uncertainties in the flux ratios, differential extinction in the lens galaxy and more 
complicated mass distributions for modeling the lens galaxy could have an important impact on the 
estimates of $\kappa$ and $\gamma$.  Therefore,} the values for $\kappa$ and $\gamma$ in 
Table \ref{table1} computed for an SIS+$\gamma_e$ model should  be taken individually only as 
compatible values with high uncertainties. However, we will assume that
the distribution of values in its entirety can be considered as statistically representative 
for the sample of observed image pairs. Fortunately, the uncertainty in the macro-lens models does 
not play a crucial role in the conclusions of our study.

\section{Statistical Analysis of the Observed Microlensing Magnifications \label{sect_stat}}

To analyze the microlensing magnification data, we need to consider that each $\Delta m$ measurement 
results from the flux ratio of two images seen through different locations at the lens galaxy. The 
microlensing magnification probability of a given image, $f_{{\kappa_*}_1,\kappa_{1},\gamma_{1}}(m_1)$, 
depends on the projected matter density in compact objects, ${\kappa_*}_1$,  the total projected mass
 density, $\kappa_{1}$, and the shear, $\gamma_1$. Thus, the probability distribution of the difference 
 in microlensing magnification of a pair of images, $\Delta m= m_1- m_2$, is given by the  integral

\begin{equation}
f_{{\kappa_*}_1,{\kappa_*}_2,\kappa_{1},\kappa_{2},\gamma_{1},\gamma_{2}}(\Delta m)=\int{f_{{\kappa_*}_1,\kappa_{1},\gamma_{1}}(m_1)f_{{\kappa_*}_2,\kappa_{2},\gamma_{2}}(m_1-\Delta m)d m_1}.
\end{equation}

To simplify the analysis we will suppose that the fraction of matter in compact objects, 
$\alpha=\kappa_*/\kappa$, is the same everywhere. The probability distribution of the 
difference in microlensing magnification of a pair of images can then be written as

\begin{equation}
f_{\alpha\kappa_{1},\alpha\kappa_{2},\kappa_{1},\kappa_{2},\gamma_{1},\gamma_{2}}(\Delta m)=\int{f_{\alpha\kappa_{1},\kappa_{1},\gamma_{1}}(m_1)f_{\alpha\kappa_{2},\kappa_{2},\gamma_{2}}(m_1-\Delta m)dm_1}.
\label{eq_convol}
\end{equation}

From this expression we can evaluate the probability of obtaining a microlensing measurement 
$\Delta m_i$ from a pair of images, 
$f^i_{\alpha\kappa^i_{1},\alpha\kappa^i_{2},\kappa^i_{1},\kappa^i_{2},\gamma^i_{1},\gamma^i_{2}}(\Delta m_i)$. 
Then, to estimate $\alpha$ using all the available information we maximize the likelihood function corresponding to the $N$ measurements collected in Table \ref{table0},

\begin{equation}
\log L(\alpha)=\sum_{i=1}^N{\log f^i_{\alpha\kappa^i_{1},\alpha\kappa^i_{2},\kappa^i_{1},\kappa^i_{2},\gamma^i_{1},\gamma^i_{2}}(\Delta m_i)}.
\label{eq_mle}
\end{equation}

\subsection{Probability Distributions of the Difference in Microlensing 
Magnifications for Image Pairs, $f_{\alpha\kappa_{1},\alpha\kappa_{2},\kappa_{1},\kappa_{2},\gamma_{1},\gamma_{2}}$ \label{sect_maps}}

We first compute the microlensing magnification probability distributions for one image,  
$f_{\alpha\kappa,\kappa,\gamma}(m)$. The first step is to simulate microlensing magnification 
maps for the different values of $\kappa$ and $\gamma$ in Table \ref{table1}. We consider 
several values for the fraction of mass in compact objects\footnote{ This sequence of 
microlensing maps parameterized by $\alpha=\kappa_*/\kappa$ assumes that the overall mass distribution 
(compact objects and smooth mass distribution) is close to isothermal. However, in many studies 
(e.g. \citealp{2009arXiv0906.4342D}) the lens galaxy is simulated with a constant mass-to-light (M/L) 
ratio model representing the galaxy stellar content (typically a de Vaucouleurs profile) embedded in a 
smooth halo of dark matter with no compact objects (a NFW halo, for instance; \citealp{1996ApJ...462..563N}). 
In this case, the sequence of models is parameterized by $f_{M/L}$, the fraction of mass in the 
stellar component relative to a constant M/L ratio model with no halo (that is, the model with $f_{M/L}=1$). 
Although the meanings of $\alpha$ and $f_{M/L}$ are different, the results of both procedures can be 
compared obtaining from each $f_{M/L}$ model values of $\kappa$ and $\kappa_*$.}: 
$\alpha=1,  0.5, 0.3, 0.25, 0.2, 0.15, 0.10, 0.05, 0.03$, and $0.01$. The histogram of 
each magnification map then provides a frequency distribution model of microlensing magnifications.  

We obtain square maps 24 Einstein radii on a side with a spatial resolution of 0.012 Einstein 
radii per pixel. To compute the magnification maps we use the inverse polygon mapping method 
described in \citet{2006ApJ...653..942M}. An example of the maps is shown in Figure \ref{fig_maps}.  
The microlensing magnification at a given pixel is then obtained as the ratio of the magnification in 
the pixel to the average magnification. Histograms of these normalized maps give the relative frequency 
of microlensing magnifications for a pixel-size source (see some examples in Figure \ref{fig_pdf}). 
These distributions are in agreement with the results obtained with a different method by \citet{1995MNRAS.276..103L}. 

To model the unresolved quasar source we consider a Gaussian with $r_s=2.6 \cdot 10^{15}\rm\, cm$ (1 light-day) 
\citep{2002ApJ...579..127S,2004ApJ...605...58K}. The convolution of this Gaussian with the ``pixel'' maps gives 
the magnification maps for the quasar. For a system with redshifts $z_l\sim 0.5$ and $z_s\sim 2$ for the lens 
and the source respectively, the Einstein radius for a compact object of mass $M$ is 
$\eta_0\sim 5.2 \cdot 10^{16}\sqrt{M/M_\odot}\,\rm cm$. Thus for $M=1M_\odot$, $\eta_0\sim 5.2 \cdot 10^{16}\rm cm$, 
and the size of a pixel is $6.2 \cdot 10^{14}\rm cm$.

Finally, the histograms of the convolved maps give the frequency distributions of microlensing 
magnifications, $f_{\alpha\kappa,\kappa,\gamma}(m)$, which show differences with respect to the 
results obtained for a pixel-size source at the high magnification wing (the same effect that can be 
observed in \citealp{1995MNRAS.276..103L}). From the  cross-correlation of pairs of these individual 
probability functions $f_{\alpha\kappa_1,\kappa_1,\gamma_1}(m_1)$, and $f_{\alpha\kappa_2,\kappa_2,\gamma_2}(m_2)$ (see 
Eq. \ref{eq_convol}) we obtain the probability function of the difference in microlensing magnification between two 
images, $f_{\alpha\kappa_{1},\alpha\kappa_{2},\kappa_{1},\kappa_{2},\gamma_{1},\gamma_{2}}(\Delta m=m_1-m_2)$.  In
 Figures \ref{fig_models_tot_D}, \ref{fig_models_tot_E}, and \ref{fig_models_tot_F} the 
 $f_{\alpha\kappa_{1},\alpha\kappa_{2},\kappa_{1},\kappa_{2},\gamma_{1},\gamma_{2}}(\Delta m)$ 
                            distributions corresponding to the 29 image pairs of Table \ref{table1} are plotted.

\subsection{Maximum Likelihood Estimate of the Fraction of Mass in Compact Objects, $\alpha$. 
Confidence Intervals \label{sect_mle}}

{In Figure \ref{fig_mle_tot_B} we present $\log L(\alpha)$  (see Eq. \ref{eq_mle}). Using the 
$\log L(\alpha\pm n\sigma_\alpha)\sim \log L_{max}-n^2/2$ criterion we derive $\alpha(\log L_{max})=0.10_{-0.03}^{+0.04}$ (90\% confidence interval).}

The maximum likelihood method can be affected by errors in the microlensing measurements, 
$\sigma_{\Delta m_i}$. From eq. \ref{eq_mle} we obtain,

\begin{equation}
\Delta \log L(\alpha)=\sum_{i=1}^N{{1 \over f^i_{\alpha\kappa^i_{1},\alpha\kappa^i_{2},\kappa^i_{1},\kappa^i_{2},\gamma^i_{1},\gamma^i_{2}}(\Delta m_i)} {\partial f^i_{\alpha\kappa^i_{1},\alpha\kappa^i_{2},\kappa^i_{1},\kappa^i_{2},\gamma^i_{1},\gamma^i_{2}}(\Delta m_i)\over \partial \Delta m_i}\sigma_{\Delta m_i}}.
\end{equation}

According to this last expression, microlensing measurement errors do not significantly affect  
the likelihood of flat probability distributions (typical of large values of $\alpha$). On the 
contrary, the likelihood functions corresponding to low values of $\alpha$ (associated with sharply
 peaked probability distributions) can be strongly modified by the microlensing measurement errors.
  Notice, moreover, that these changes tend to penalize the low $\alpha$ hypothesis.

To show the impact of $\sigma_{\Delta m_i}$ on the maximum likelihood estimate of $\alpha$, in 
Figure \ref{fig_mle_tot_B}  we {also} present $\log L(\alpha)$  (see Eq. \ref{eq_mle}) with error 
bars, $\pm \Delta \log L(\alpha)$, estimated considering that each $\Delta m_i$ is a normally 
distributed variable with $\sigma_{\Delta m_i}=0.20$ (a realistic estimate). Using the
 $\log L(\alpha\pm n\sigma_\alpha)\sim \log L_{max}-n^2/2$ criterion and taking into account 
 the error bars of $\log L(\alpha)$, we derive $\alpha(\log L_{max})=0.05_{-0.03}^{+0.09}$ (90\% confidence interval).

\subsection{Influence of the Continuum Source Size. Influence of the Microlenses Mass \label{sect_scaling}}

Increasing the {size parameter of the Gaussian representing the} continuum source, $r_s$, affects
 the previous results by smoothing the magnification patterns and, consequently, the probability 
 distributions. {To study the dependence of the estimate of $\alpha$ on the source size we have 
 computed probability and likelihood functions for several values of  this parameter, 
 $r_s=0.62 \cdot 10^{15},\, 2.6 \cdot 10^{15},\, 8 \cdot 10^{15}$, and $26 \cdot 10^{15}\rm\, cm$.} 
 {To correct $r_s$ from projection effects we have taken into account that the intrinsic and
  projected source areas are related by a $\cos i$ factor; that is $r_s\sim \sqrt{\cos i}\, r_{s_0}$. 
  Assuming that the (disk) sources are randomly oriented in space (the probability of finding a disk with 
  inclination, $i$, proportional to $\sin i$) and averaging on the inclination, we obtain 
  $r_{s_0}\sim 1.5 r_s$. In Figure \ref{fig_scaling} we present the likelihood functions corresponding to 
  sources of several deprojected size parameters, $r_{s_0}$. In Figure \ref{fig_alfa_scaling} we plot the 
  maximum likelihood estimate of $\alpha$ versus\footnote{Note that for the considered Gaussian 
  intensity profile the radii enclosing 50\% and 90\% of the source energy are related to the Gaussian 
  source size parameter, $r_{s_0}$, according to $r_{1/2}=r(50\%)=1.18r_{s_0}$ and $r(90\%)=2.1r_{s_0}$} 
  $r_{s_0}$. Error bars correspond to 90\% confidence intervals. According to this figure, low values of 
  $\alpha$ are expected for continuum source sizes, $r_{s_0}$, of order $10^{16}\rm\, cm$ { or less. 
  Observing microlensing variability for nine gravitationally lensed quasars \citet{2007arXiv0707.0305M} 
  measure the accretion disk size. The average value of the nine half-light radius determinations is  
  $<r_{1/2}>=6 \cdot 10^{15}\rm\, cm$. For this value we found (see Figure \ref{fig_alfa_scaling}) 
  $\alpha= 0.05_{-0.03}^{+0.09}$. \citet{2007arXiv0707.0305M} report a scaling between the accretion disk 
  size and the black hole mass. In the range of black hole masses considered by \citet{2007arXiv0707.0305M} 
  the maximum is $M_{BH}=2.37 \cdot 10^9 M_\odot$, which,  using the scaling derived by these authors, 
  corresponds to $r_{1/2}=2.4 \cdot 10^{16}\rm\, cm$. For this size we obtain (see 
  Figure \ref{fig_alfa_scaling}) $\alpha\sim 0.10$. Values 
  $M_{BH}\ge  10^{10} M_\odot$ ($r_{1/2}\ge 3.4 \cdot 10^{16}\rm\, cm$) should be considered to obtain $\alpha\ge 0.20$. 
  On the other hand, \citet{2007ApJ...661...19P} comparing X-ray and optical microlensing in a 
  sample of ten lensed quasars inferred $r_{1/2}\sim 1.3 \cdot 10^{16}\rm\, cm$. For this size we 
  obtain (see Figure \ref{fig_alfa_scaling}) $\alpha=0.10_{-0.06}^{+0.05}$. Thus, according to these 
  recent size estimates based on the observations of two relatively large samples of gravitational lenses,} 
  high values of $\alpha$ are possible only if the continuum source size is substantially larger than expected.

Owing to the scaling of the Einstein radius with mass, $\eta_0\propto \sqrt{M}$, a change in the 
mass of microlenses can be alternatively seen as a change in the spatial scaling of the magnification 
pattern that leaves invariant the projected mass density, $\kappa$. Thus, multiplying the mass of the 
microlenses by a factor $C$ (and leaving unaltered the continuum size) is equivalent to multiplying 
the size of the continuum source by a factor $1/\sqrt{C}$ (leaving unaltered the masses of microlenses). 
Then the {computed} models corresponding to sources of sizes $r_s=0.62 \cdot 10^{15},\, 2.6 \cdot 10^{15}$ 
and $8 \cdot 10^{15} \,\rm cm$ (with $1 M_\odot$ microlenses) are equivalent to models corresponding to 
microlens masses of $17$, $1$, and $0.1M_\odot$ (with $r_s=2.6 \cdot 10^{15}\rm\, cm$).  This result 
implies that the probability models do not differ significantly if we change the mass of microlenses 
between 17 and 0.1$M_\odot$  Thus, microlensing statistics are insensitive to changes of mass in the 
expected range of stellar masses.

\section{Discussion and Conclusions \label{sect_conclusions}}

In the previous sections we have extended to the extragalactic domain the local (Milky Way, LMC, and M31) 
use of microlensing to probe the properties of the galaxy halos. Although our primary aim was to explore 
the practical application of the proposed method, we found that the data available in the literature can 
be consistently interpreted only under  the hypothesis of a very low mass fraction in microlenses; 
{ at 90\% confidence: $\alpha(\log L_{max})=0.05_{-0.03}^{+0.09}$ (maximum likelihood estimate)}  {for a 
quasar continuum source of intrinsic size  $r_s=2.6 \cdot 10^{15}\rm\, cm$}. This result arises directly 
from the shape of the histogram of microlensing magnifications, with a maximum of events close to no 
magnification and stands for a wide variety of microlensing models statistically representative of the considered 
image pairs. There is a dependence of the estimate of $\alpha$ on the source size but high values of the 
mass fraction ($\alpha>0.2$) are possible only for unexpectedly large source sizes ($r_s>4 \cdot 10^{16}\rm\, cm$). 
The low mass fraction is in good agreement with the results of EROS \citep{{2007A&A...469..387T}} for the 
Milky Way and with the limit established by MEGA \citep{2006A&A...446..855D} for M31. { The agreement is 
also good with the few microlensing-based estimates available for individual objects. In RXJ 1131-1231 
\citet{2009arXiv0906.4342D} found $\alpha\sim 0.1$. In PG 1115+080 \citet{2008ApJ...689..755M} obtained 
values in the range $\alpha=0.08$ to $0.15$. For the same system \citet{2009ApJ...697.1892P} found  
$\alpha\sim 0.1$ for a source of size $r_s=1.3 \cdot 10^{16}\rm\, cm$}. 

{On the other hand, our estimate of the fraction of mass in microlenses, $\alpha(\log L_{max})=0.05_{-0.03}^{+0.09}$, 
approximates the expectations for the fraction of visible matter. {  \cite{{2007ApJ...671.1568J}}, for instance, 
comparing the mass inside the Einstein ring in 22 gravitational lens galaxies with the mass needed to produce 
the observed velocity dispersion, inferred average stellar mass fractions of $0.026\pm 0.006$ (neglecting adiabatic 
compression) and $0.056\pm 0.011$ (including adiabatic compression). As discussed in \cite{{2007ApJ...671.1568J}} 
these values are also in agreement with other estimates of the stellar mass fraction that relied on stellar 
population models: $\sim 0.08$ \citep{2006ApJ...648..826L},  $0.065_{-0.008}^{+0.010}$ \citep{2005ApJ...635...73H}, 
and $0.03_{-0.01}^{+0.02}$ \citep{2006MNRAS.368..715M}. Thus, we can conclude} that microlensing is { probably} 
caused by stars in the lens galaxy, and that there is no statistical evidence for MACHOS in the halos of the 20 
galaxies of the sample we considered.

How robust are these results? There are several sources of uncertainty to consider.  Firstly, we neglected in 
Equation \ref{eq_baseline} the term arising from source variability, $\Delta m_0^{con}$. From  a group of 17 
gravitational lenses with photometric monitoring available in the literature we estimate an average gradient of 
variability of $0.1\rm\, mag\, year^{-1}$. Taking into account that the average delay between images is about 
$3\rm\, months$ (a conservative estimate; note that the group of  lens systems used includes many doubles, 
some of them with very large time delays) we can expect an amplitude related to intrinsic source variability of 
$\Delta m_0^{con}\sim 0.03$, which, according to the histogram of magnifications (Figure \ref{fig_histo_A}), is not 
significant. Moreover, if we assume that the probability of $\Delta m_0^{con}$ is normally distributed, the global 
effect of source variability is to broaden the histogram of microlensing magnifications, diminishing the peak and 
enhancing the wing. In other words, source variability leads to an overestimate of $\alpha$. Thus, the mass 
fraction should be even lower if significant source variability were hidden in the data. In the same way, other 
sources of error in the data, such as the difficulty in separating line and continuum or in removing from the narrow 
emission lines the high ionization broad emission lines that could be partially affected by microlensing, probably 
tend to induce additional magnitude differences, $\Delta m$, between the images and, hence, to an overestimate of 
$\alpha$. On the contrary, cross-contamination between the spectra of a pair of images masks the impact of
 microlensing and may affect our results. Although most of the bibliographic sources of microlensing measurements 
 analyze this problem concluding that the spatial resolution was sufficient to extract the spectra without 
 contamination, it is clear that high S/N data obtained in subarcsecond seeing conditions will help to control this 
 important issue. 

Another point to address is the treatment of some of the quads, where only a subset of the images are used. Are we 
systematically excluding faint images that might be highly demagnified by microlensing? Let us examine the four 
incomplete quads in our sample. The fold lens SDSS J1004+4112 has two close images A and B. A is probably a 
saddle-point image and shows the most anomalous flux (Ota et al. 2006). In contrast, the optical/X-ray flux 
ratios of C and D are almost the same. Thus, there is no reason to suppose that the image without a useful 
spectrum (D) has higher microlensing probability than the others. PG 1115+080 is another fold quad. $A_1$ and $A_2$ 
are the two images closest to the critical curve and have a (moderately) anomalous flux ratio and optical variability  
(Pooley et al. 2007). The two images without available spectra (C and D) show only a  small optical variability 
and are not particularly prone to microlensing. In RXS J1131-1231 the most anomalous flux ratio is B/C and A is a 
saddle-point image (Sluse et al. 2006). Image D (the one with no available spectrum)  also has an anomalous flux but 
is not more susceptible to microlensing than the other images. Thus, in three of the four incomplete quads there is 
no reason to suppose that we are biasing the sample towards image pairs with lower microlensing probability. The 
case of SDSS 0924+0219 is more problematic. There are two sets of data for this  object, one by Eigenbrod et 
al. (2006) based on observations of the low ionization lines MgII and CIII], which, after two epochs of observation, 
reveals no difference between the line and continuum flux ratios of components A and B. The other set of data 
(Keeton et al. 2006) is based on Ly$\alpha$ observations (a high ionization emission line suposed to come from a 
smaller region than the low ionization emission lines) and microlensing is detected not only in the continuum but 
also in the emission lines. This implies that the baseline for no microlensing magnification cannot be defined and, 
consequently, we could not  consider Keeton et al. (2006) results. Anyway, we have repeated (as a test)  the entire
maximum likelihood estimate procedure  to  derive $\alpha$ but now using for SDSS 0924+0219 the microlensing measurements 
by Keeton et al. (2006). The results are almost identical: $\alpha = 0.05^{+0.10}_{-0.03}$. 

The size of the sample also limits the statistical interpretation. An improvement in the S/N of the histogram of 
microlensing magnifications is very important to ascertain the statistical significance of the low frequency of 
events at large magnification (only two events of high magnification are detected), which can impose severe constraints 
on the microlensing models. Another reason to increase the size of the sample is the possibility to define subsamples 
at different galactocentric distances where different ratios of visible to dark matter are expected. In the same way it 
would be possible to define subsamples according to the type of lens galaxy or other interesting properties of lens systems. 

In any case, the impact of the main result of our study $-$absence of MACHOS in the 10 to 0.1$M_\odot$ mass range in the 
halos of lens galaxies$-$ and its future prospects, points to the need to improve the statistical analysis in two ways: 
increasing the number, quality and homogeneity of the microlensing magnification measurements from new observations, 
and reducing the uncertainties in the macro-lens models.

\acknowledgments

We thanks the anonymous referee for valuable comments and suggestions. We are grateful to A. Eigenbrod, P.Green, M. Oguri, L. Wisotzki and O. Wucknitz for kindly providing spectra. This work was supported by the European Community's Sixth Framework Marie Curie RTN (MRTN-CT-505183 "ANGLES") and by the Ministerio of Educaci\'on y Ciencia of Spain with the grants AYA2004-08243-C03-01 and AYA2004-08243-C03-03. V.M. acknowledges support by FONDECYT grant 1071008. J.A.M. is also supported by the Generalitat Valenciana with the grant PROMETEO/2009/64.

\clearpage

\begin{figure}[h]
\plotone{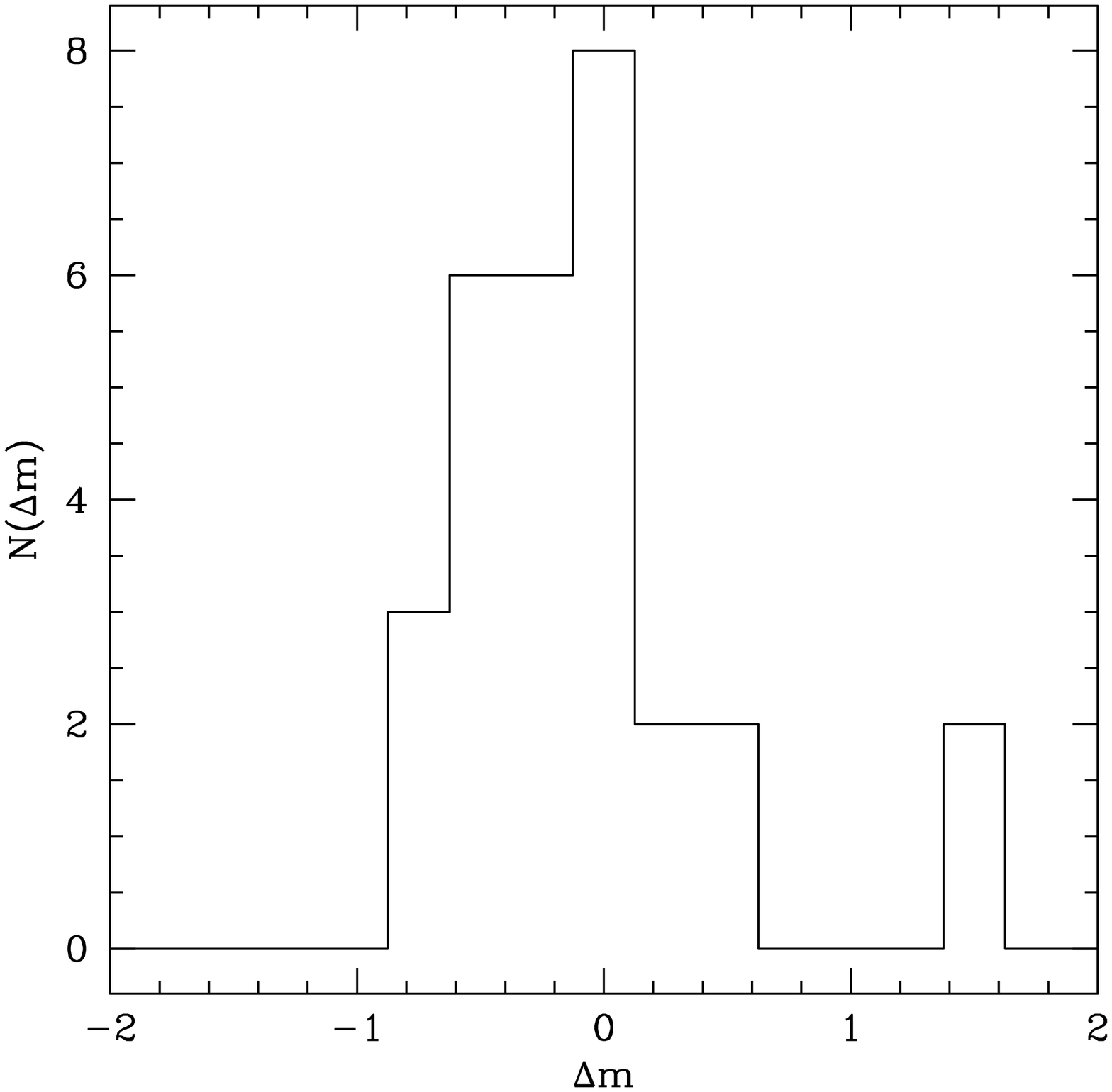}
\caption{Histogram of microlensing magnifications for the sample of image pairs in Table \ref{table1} (bin=0.25). \label{fig_histo_A}}
\end{figure}


\begin{figure}[h]
\plotone{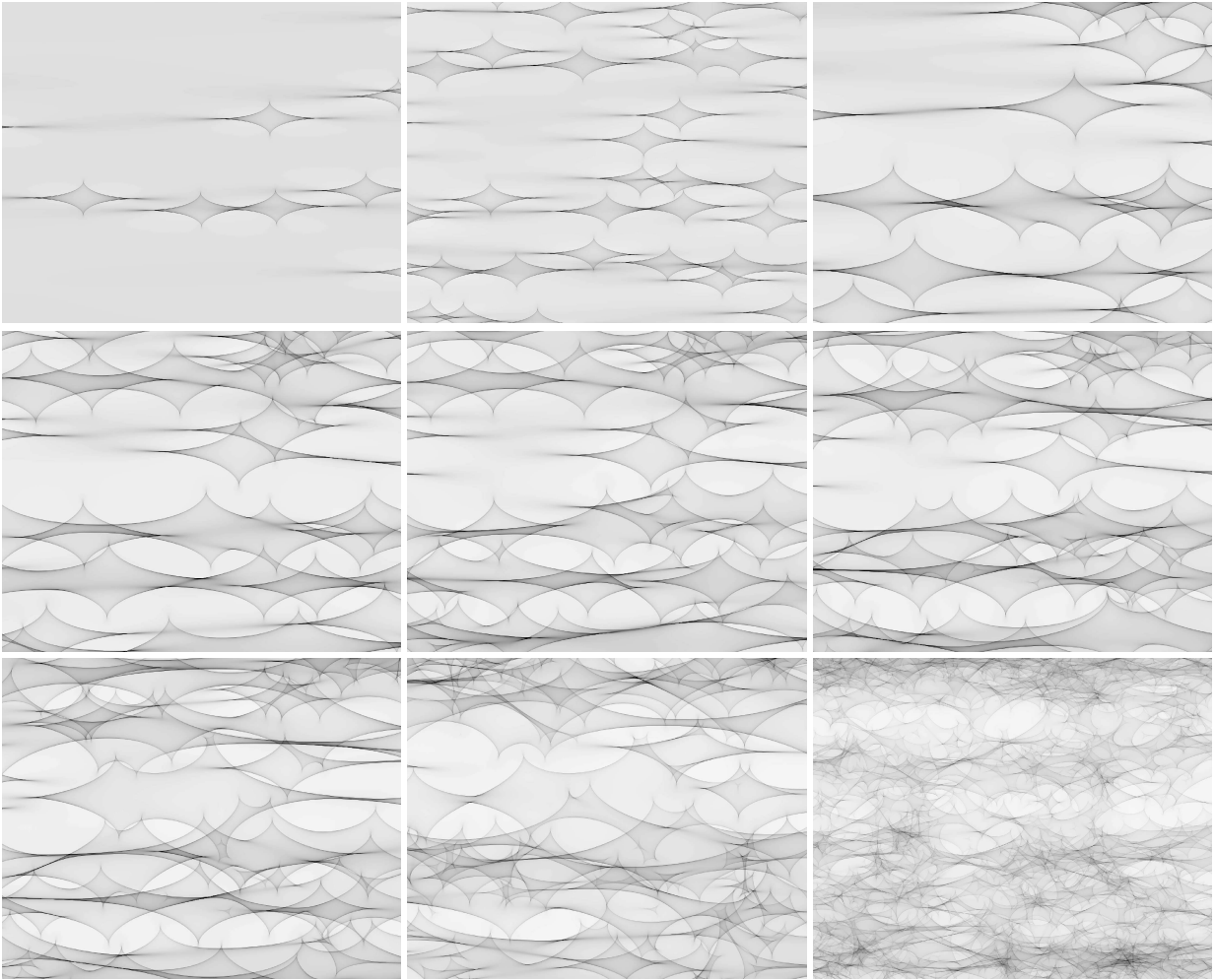}
\caption{ Example of magnification maps for the case $\kappa=\gamma=0.45$. From top to bottom and from left to right, 
maps correspond to $\alpha=$ 0.01, 0.05, 0.10, 0.15, 0.20, 0.25, 0.30, 0.50, 1.00. \label{fig_maps}}
\end{figure}

\begin{figure}[h]
\plotone{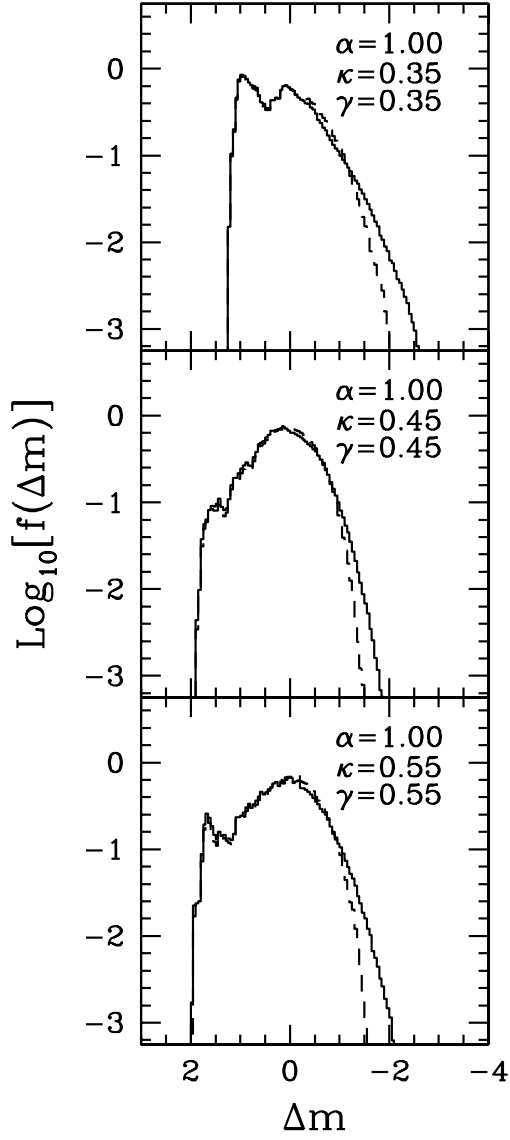}
\caption{ Relative frequency of microlensing magnifications, $f_{\alpha\kappa,\kappa,\gamma}$, for pixel size 
(solid line) and $r_s=2.6 \cdot 10^{15}\rm\, cm$ (dashed line) sources (see text). Examples for three different values
 of $\kappa$ in the case $\kappa=\gamma$, see text. \label{fig_pdf}}
\end{figure}

\begin{figure}[h]
\plotone{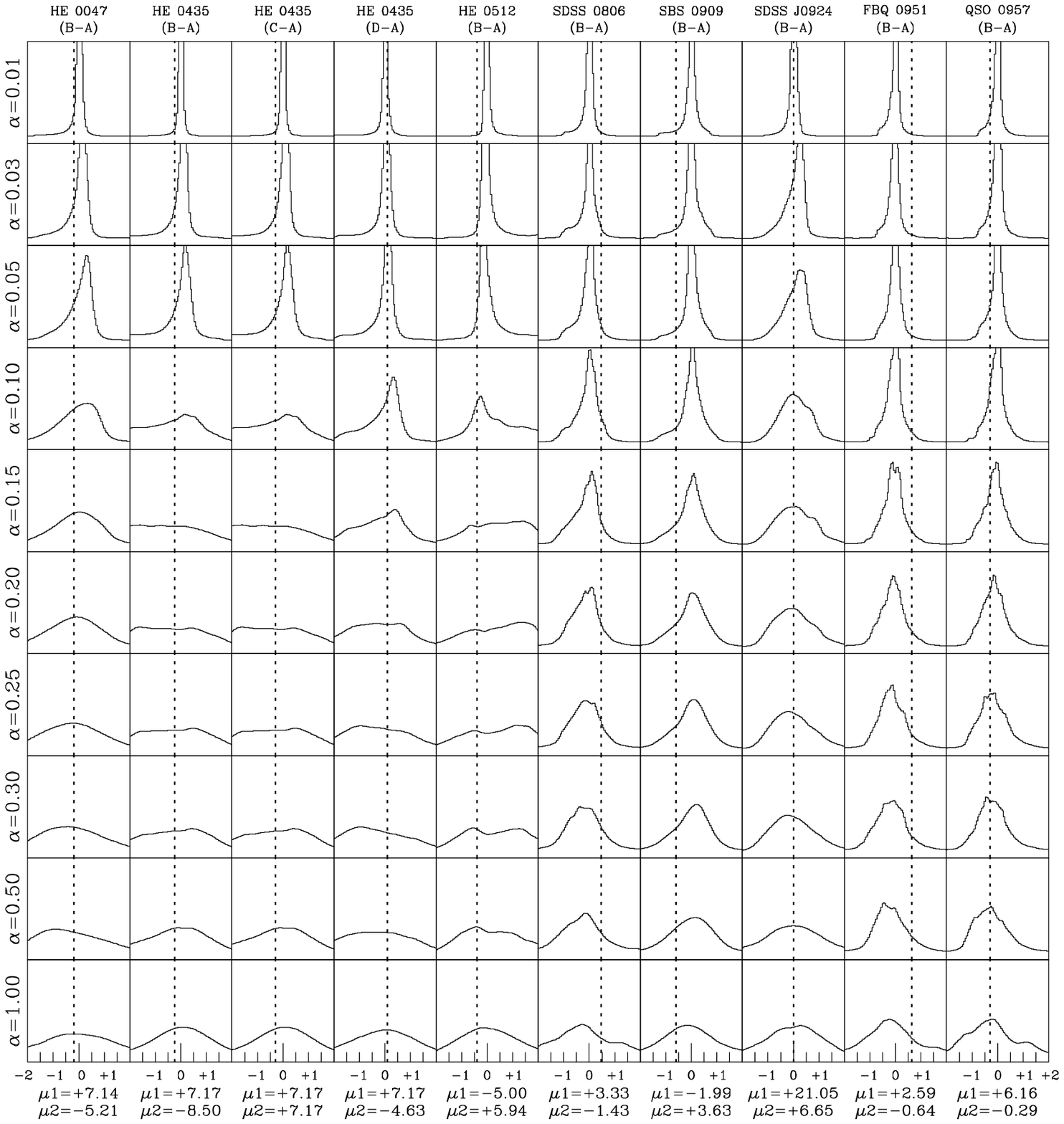}
\caption{Probability models, $f_{\alpha\kappa_{1},\alpha\kappa_{2},\kappa_{1},\kappa_{2},\gamma_{1},\gamma_{2}}(\Delta m=m_1-m_2)$, 
corresponding to each image pair in the sample for different values of the fraction of mass in compact objects, $\alpha$ (see text). 
$\mu_1$ and $\mu_2$ are the magnifications of the images considered in each pair. The vertical dashed line corresponds to the 
microlensing measurement value.\label{fig_models_tot_D}}
\end{figure}

\begin{figure}[h]
\plotone{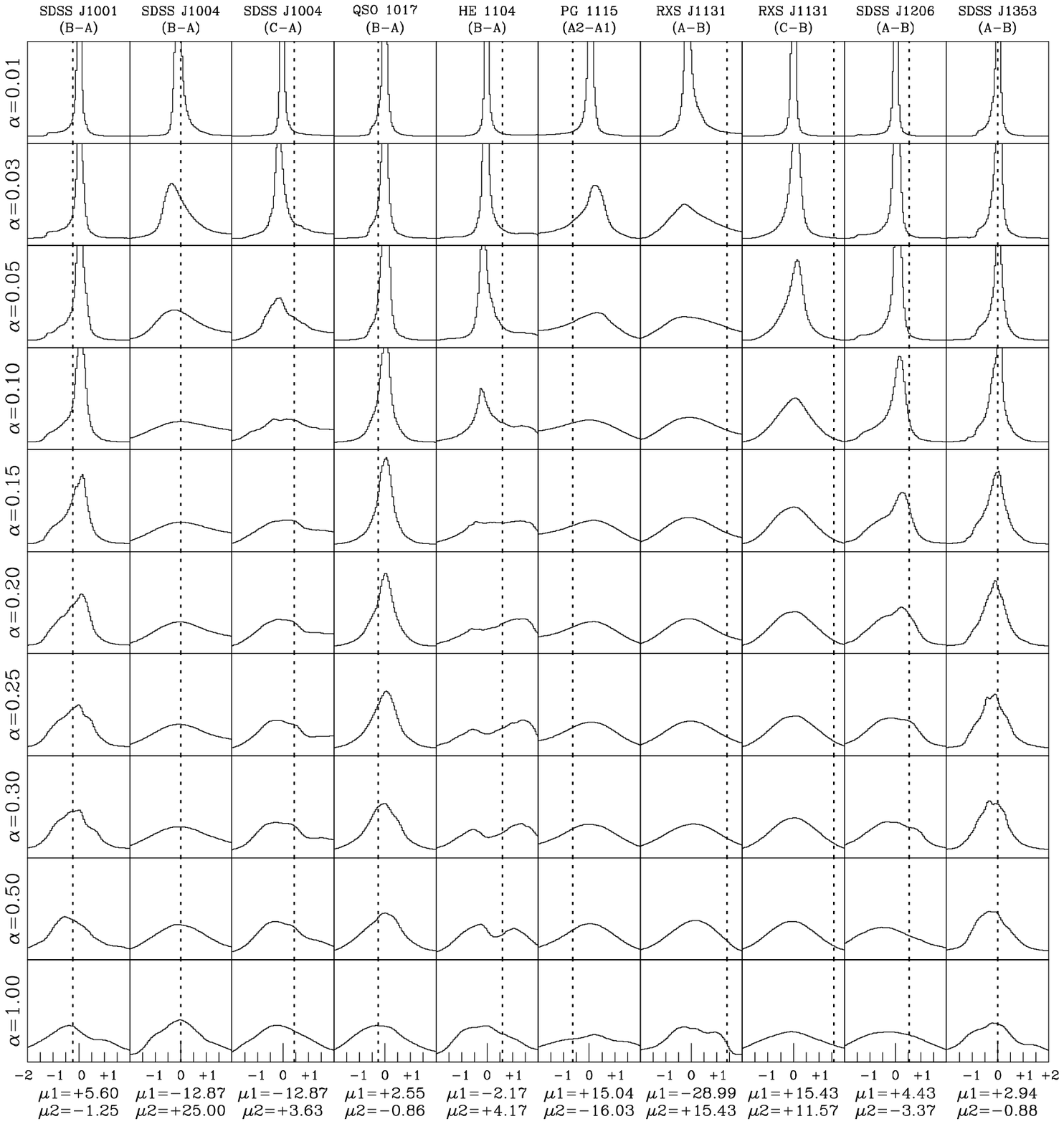}
\caption{Probability models, $f_{\alpha\kappa_{1},\alpha\kappa_{2},\kappa_{1},\kappa_{2},\gamma_{1},\gamma_{2}}(\Delta m=m_1-m_2)$, 
corresponding to each image pair in the sample for different values of the fraction of mass in compact objects, $\alpha$ (see text). 
$\mu_1$ and $\mu_2$ are the magnifications of the images considered in each pair. The vertical dashed line corresponds to the 
microlensing measurement value. \label{fig_models_tot_E}}
\end{figure}

\begin{figure}[h]
\plotone{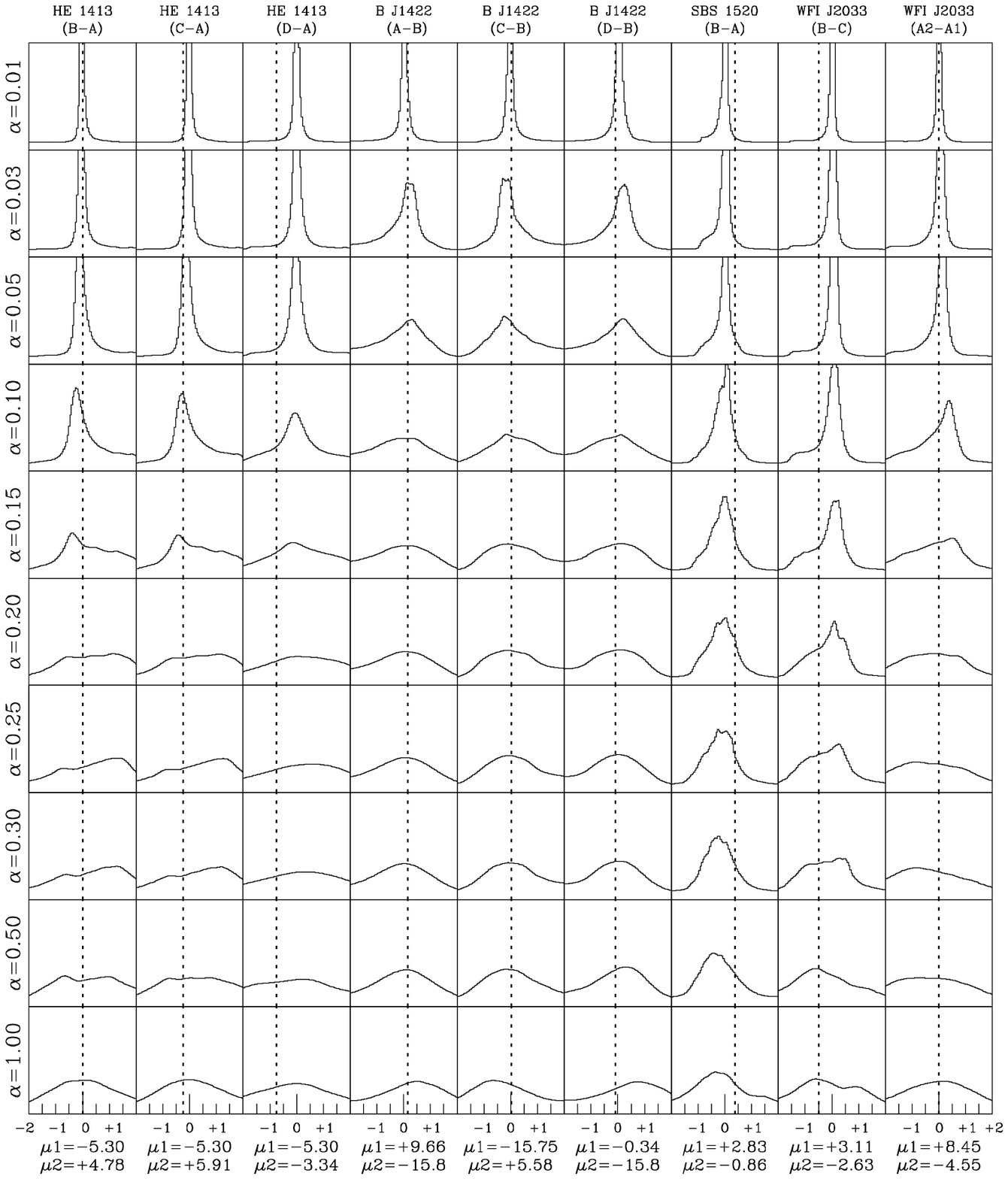}
\caption{Probability models, $f_{\alpha\kappa_{1},\alpha\kappa_{2},\kappa_{1},\kappa_{2},\gamma_{1},\gamma_{2}}(\Delta m=m_1-m_2)$,
 corresponding to each image pair in the sample for different values of the fraction of mass in compact objects, $\alpha$ (see text). 
 $\mu_1$ and $\mu_2$ are the magnifications of the images considered in each pair. The vertical dashed line corresponds to the
  microlensing measurement value. \label{fig_models_tot_F}}
\end{figure}

\begin{figure}[h]
\plotone{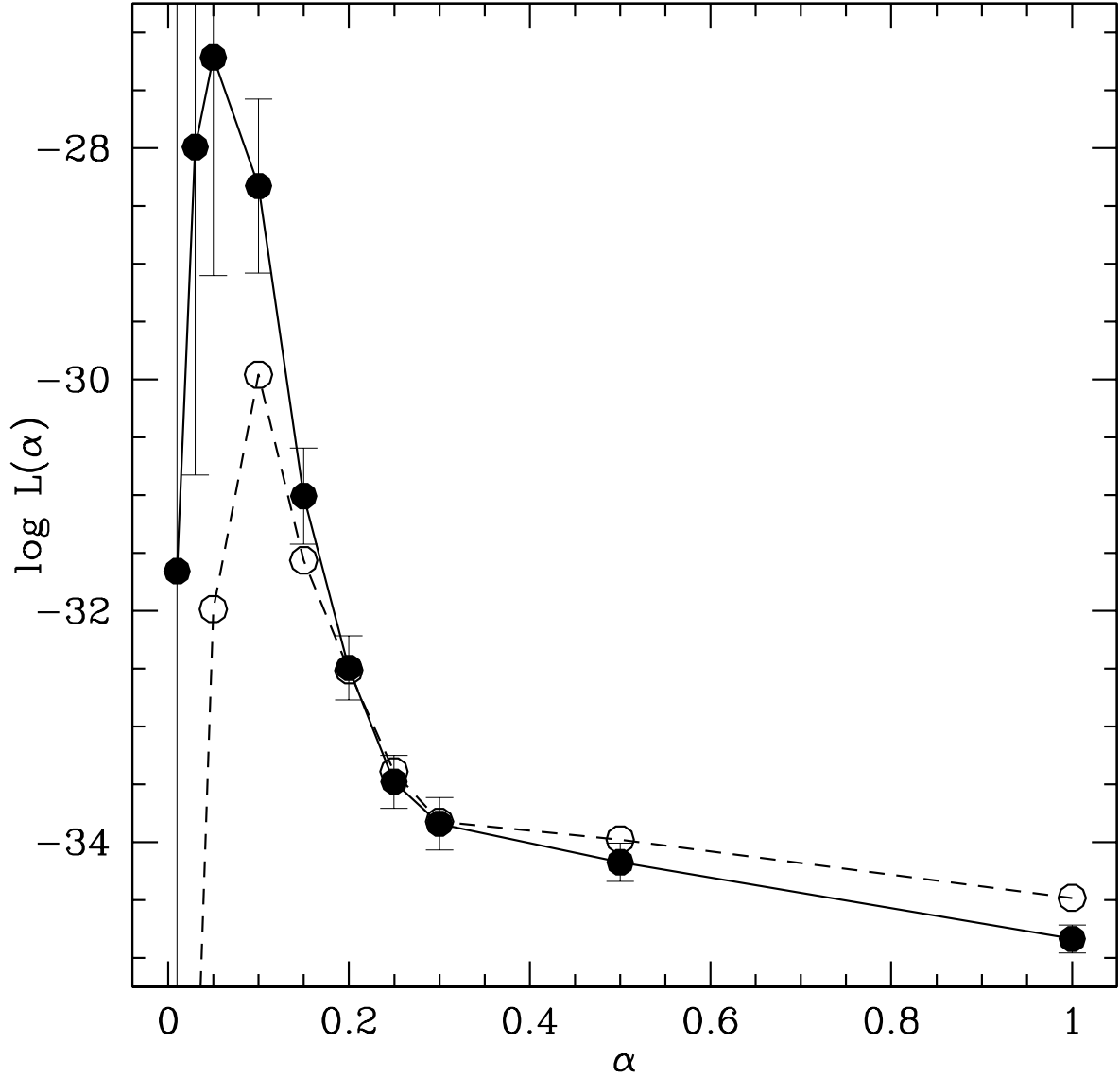}
\figcaption[]{Likelihood function vs. fraction of mass in compact objects. Dashed line (circles): likelihood function 
supposing that the microlensing measurements are unaffected by errors. Continuous line (filled circles): likelihood 
function supposing that the microlensing measurements are affected by 0.2 mag errors.  See text for details. \label{fig_mle_tot_B}}
\end{figure}

\begin{figure}[h]
\plotone{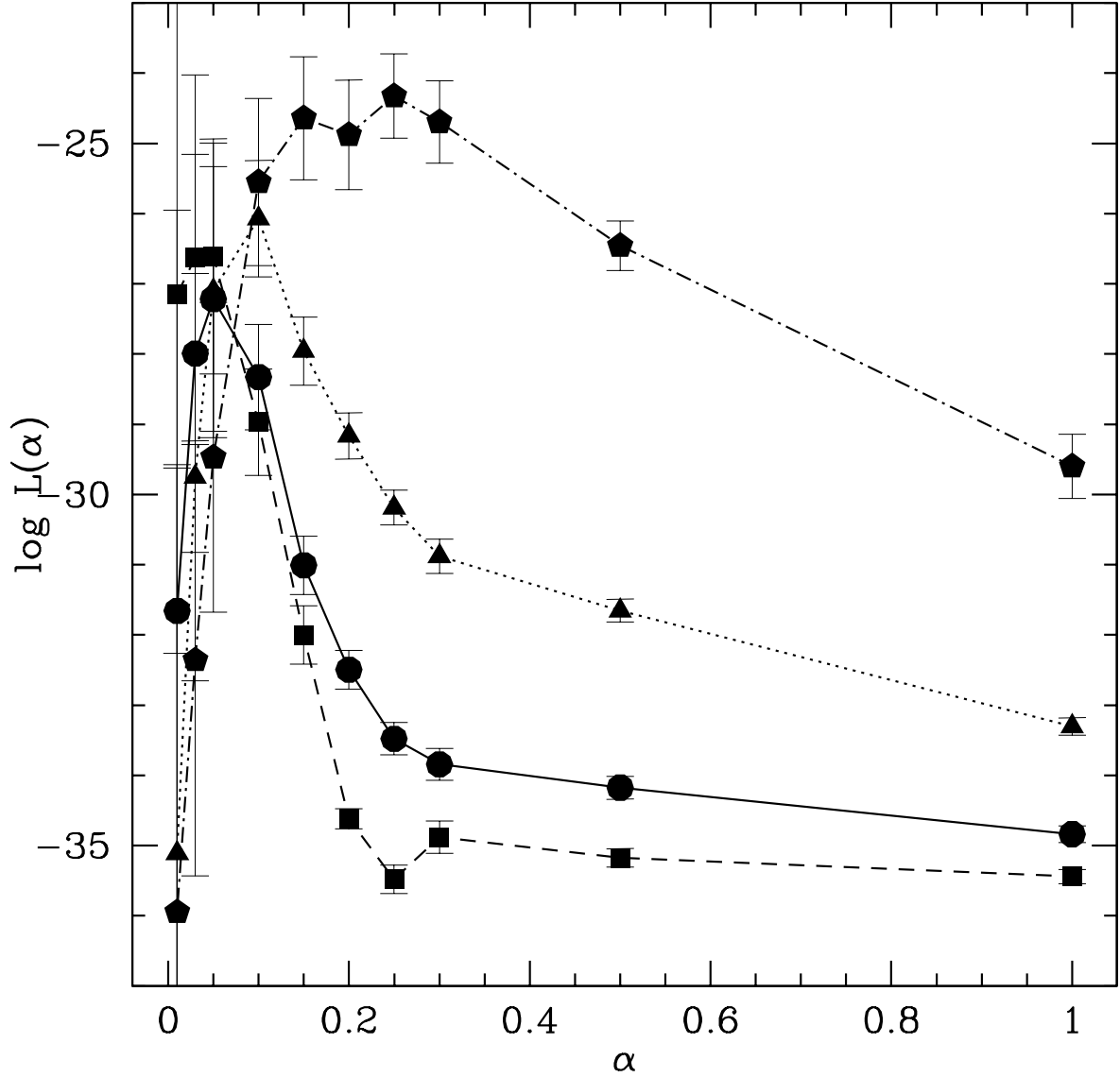}
\figcaption[]{Likelihood functions corresponding to sources of deprojected size parameter, $r_{s_0}$: $0.8 \cdot 10^{15}\rm\, cm$ 
(squares; dashed line), $3.9 \cdot 10^{15}\rm\, cm$ (circles; continuous line), $12.4 \cdot 10^{15}\rm\, cm$ (triangles; 
dotted line), and $40.4 \cdot 10^{15}\rm\, cm$ (pentagons; dot-dashed line). \label{fig_scaling}}
\end{figure}

\begin{figure}[h]
\plotone{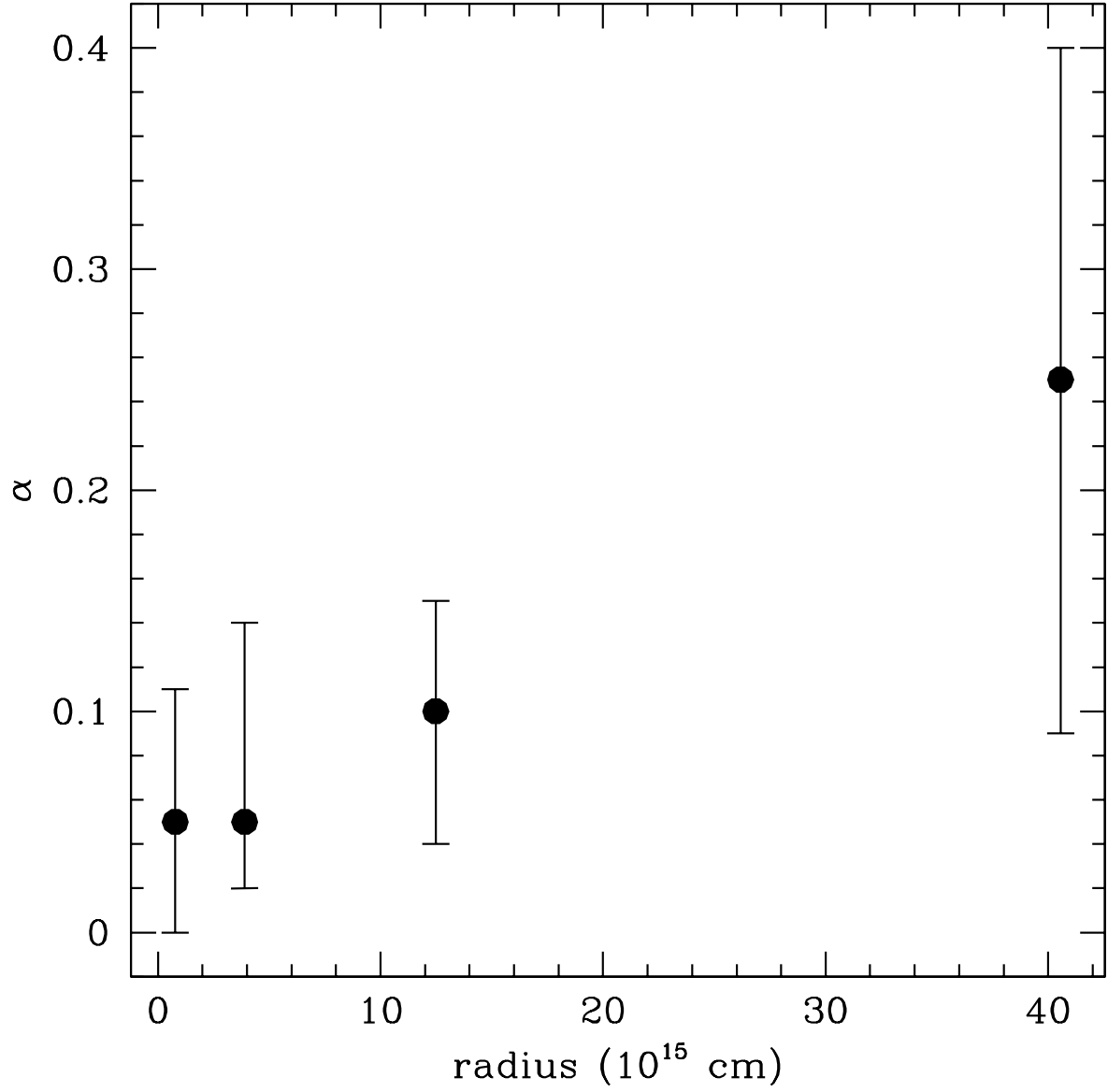}
\figcaption[]{Maximum likelihood estimates of $\alpha$ versus deprojected source size parameter, $r_{s_0}$. Error bars 
correspond to 90\% confidence intervals. \label{fig_alfa_scaling}}
\end{figure}

\clearpage

\begin{deluxetable}{llclllllll}
\setlength{\tabcolsep}{0.06in}
\tabletypesize{\scriptsize}
\tablecaption{ \label{table0} Microlensing Data}
\tablewidth{0pt}
\tablehead{
\colhead{Object}& \colhead{Image}& \colhead{$<\Delta m>$\tablenotemark{***}}&\colhead{$\Delta m^{cont,lines}$  }&\colhead{Ly$\alpha$}&\colhead{SiIV]/OIV]}&\colhead{CIV}& \colhead{CIII]}&\colhead{MgII}&\colhead{[OIII]}\\
&\colhead{pair}&\colhead{}&cont/lines\tablenotemark{**}&cont/line\tablenotemark{*}&cont/line\tablenotemark{*}&cont/line\tablenotemark{*}&cont/line\tablenotemark{*}&cont/line\tablenotemark{*}&cont/line\tablenotemark{*}\\}
\startdata

HE 0047-1756\tablenotemark{1} &$B-A$&-0.19&$1.17/1.36$&---&---&---&---&---&---\\
HE 0435-1223\tablenotemark{2}&$B-A$&-0.24&---&---&---&---&---&---&---\\
&$C-A$&-0.30&---&---&---&---&---&---&---\\
&$D-A$&0.09&---&---&---&---&---&---&---\\
HE 0512-3329\tablenotemark{3} &$B-A$&$-0.40\pm 0.16$&---&---&---&---&---&---&---\\
SDSS 0806+2006\tablenotemark{4} &$B-A$&$-0.47\pm 0.20$&---&---&---&---&0.06/0.33&0.06/0.72&---\\
SBS 0909+532\tablenotemark{5} &$B-A$&$-0.60\pm 0.15$&---&---&---&---&---&---&---\\
SDSS J0924+0219\tablenotemark{6} &$B-A$&0.00&---&---&---&---&---&---&---\\
FBQ 0951+2635\tablenotemark{7} &$B-A$&$-0.69\pm 0.35$&---&---&---&---&0.08/1.12&0.13/0.46&---\\
QSO 0957+561\tablenotemark{8} &$B-A$&-0.30&$-0.30/0.0$&---&---&---&---&---&---\\
SDSS J1001+5027\tablenotemark{9} &$B-A$&$0.23\pm 0.04$&---&---&---&0.63/0.35&0.38/0.19&---&---\\
SDSS J1004+4112\tablenotemark{10} &$B-A$&0.00&0.50/0.50&---&---&---&---&---&---\\
&$C-A$&0.45&0.64/0.19&---&---&---&---&---&---\\
QSO 1017-207\tablenotemark{11} &$B-A$&$-0.26\pm 0.11$&---&-2.21/-2.08&-2.24/-2.06&-2.24/-1.41&-2.15/-1.76&---&---\\
HE 1104-1805\tablenotemark{12} &$B-A$&$0.60\pm 0.03$&---&---&---&---&1.75/1.12&1.68/1.12&---\\
PG 1115+080\tablenotemark{13} &$A2-A1$&-0.65&$-0.65/0.0$&---&---&---&---&---&---\\
RXS J1131-1231\tablenotemark{14} &$A-B$&1.39&---&---&---&---&---&---&0.65/-0.74\\
&$C-B$&1.58&---&---&---&---&---&---&1.27/-0.31\\
SDSS J1206+4332\tablenotemark{15} &$A-B$&$-0.56\pm 0.21$&---&---&---&---&0.32/1.08&0.54/0.89&---\\
SDSS J1353+1138\tablenotemark{16} &$A-B$&0.00&---&---&---&---&---&---&---\\
HE 1413+117\tablenotemark{17}&$B-A$&$0.00\pm 0.04$&---&---&0.23/0.19&0.20/0.23&---&---&---\\
&$C-A$&$-0.25\pm 0.10$&---&---&-0.03/0.27&-0.07/0.27&---&---&---\\
&$D-A$&$-0.75\pm 0.08$&---&---&0.2/-1.07&0.22/-0.85&---&---&---\\
B J1422+231\tablenotemark{18} &$A-B$&0.16&---&$0.27/0.11$&---&---&---&---&---\\
&$C-B$&0.02&---&$0.75/0.77$&---&---&---&---&---\\
&$D-B$&-0.08&---&$3.92/4.00$&---&---&---&---&---\\
SBS 1520+530\tablenotemark{19} &$B-A$&$-0.39\pm 0.07$&---&---&---&-0.04/0.27&0.08/0.54&---&---\\
WFI J2033-4723\tablenotemark{20} &$B-C$&-0.50&---&---&---&-0.09/0.41&---&---&---\\
&$A2-A1$&0.00&---&---&---&0.32/0.32&---&---&---\\
\enddata

\tablenotetext{*}{Magnitude differences between images in the continuum and in the line emission, respectively (when an individual value for one or more lines is available)}
\tablenotetext{**}{Magnitude differences between images in the continuum and in the line emission, respectively (when a global value for an 
spectral region including several lines is given)}
\tablenotetext{***}{Average microlensing magnification, $<\Delta m>=<\Delta m^{cont}-\Delta m^{line}>$}
\tablenotetext{1}{\cite{2004A&A...419L..31W} (Line flux ratio given by the authors. Continuum flux ratio estimated from Figure 3)}
\tablenotetext{2}{\cite{2003A&A...408..455W} (Microlensing magnifications taken from Table 3)}
\tablenotetext{3}{\cite{2003A&A...405..445W} (Microlensing magnification estimated from Figure 3)}
\tablenotetext{4}{\cite{2006AJ....131.1934I} (Flux ratios computed from electronically digitized spectra)}
\tablenotetext{5}{\cite{2005ApJ...619..749M} (Microlensing magnification estimated from Figure 7)}
\tablenotetext{6}{\cite{2006A&A...451..747E} (See text)}
\tablenotetext{7}{\cite{1998AJ....115.1371S} (Flux ratios computed from electronically digitized spectra)}
\tablenotetext{8}{\cite{2005MNRAS.360L..60G} (Line flux ratio given by the authors. Continuum ratio estimated from Figure 1)}
\tablenotetext{9}{\cite{2005ApJ...622..106O} (Flux ratios computed from electronically digitized spectra)}
\tablenotetext{10}{\cite{2006ApJ...645L...5G} (Flux ratios estimated from Figures 3 and 4)}
\tablenotetext{11}{\cite{1997A&A...327L...1S} (Flux ratios computed from electronically available spectra)}
\tablenotetext{12}{\cite{1993A&A...278L..15W} (Line flux ratio given by the authors. Continuum flux ratios estimated from Figure 3)}
\tablenotetext{13}{\cite{2005MNRAS.357..135P} (Line flux ratio given by the authors. Continuum flux ratio estimated from Figure 9)}
\tablenotetext{14}{\cite{2007A&A...468..885S} (Flux ratios taken from Table 5)}
\tablenotetext{15}{\cite{2005ApJ...622..106O} (Flux ratios computed from electronically digitized spectra)}
\tablenotetext{16}{\cite{2006AJ....131.1934I} (See text)}
\tablenotetext{17}{\cite{2005MNRAS.357..135P} (Flux ratios computed from electronically available spectra)}
\tablenotetext{18}{\cite{1996ApJ...462L..53I} (Flux ratios taken from Table 3)}
\tablenotetext{19}{\cite{1997A&A...318L..67C} (Flux ratios computed from electronically digitized spectra)}
\tablenotetext{20}{\cite{2004AJ....127.2617M} (Flux ratios estimated from Figure 9)}

\end{deluxetable}
\clearpage
\begin{deluxetable}{llcc}
\tabletypesize{\scriptsize}
\tablecaption{ \label{table3} Comparison between Emission Line and mid-IR Flux Ratios}
\tablewidth{0pt}
\tablehead{
\colhead{Object}& \colhead{Image Pair}& \colhead{$\Delta m^{lines}$}&\colhead{$\Delta m^{mid-IR}$}\\}
\startdata
SDSS J1004+4112\tablenotemark{1} &$B-A$&$0.50$&$0.30$\\
&$C-A$&$0.19$&$0.50$\\
HE 1104-1805\tablenotemark{2}&$B-A$&$1.12$&$1.13\pm 0.06$\\ 
PG 1115+080\tablenotemark{3} &$A2-A1$&$0.0$&$0.08\pm 0.06$\\
HE 1413+117\tablenotemark{4}&$B-A$&$0.21\pm 0.02$&$0.19\pm 0.07$\\
&$C-A$&$0.27\pm 0.00$&$0.36\pm 0.07$\\
&$D-A$&$0.96\pm 0.11$&$0.99\pm 0.06$\\
B J1422+231\tablenotemark{3} &$A-B$&$0.11$&$0.18\pm 0.05$\\
&$C-B$&0.77&$0.61\pm 0.06$\\
\enddata

\tablenotetext{1}{Mid-IR data from \citet{2009ApJ...702..472R}}
\tablenotetext{2}{Mid-IR data from \citet{2007ApJ...660..146P}}
\tablenotetext{3}{Mid-IR data from \citet{2005ApJ...627...53C}}
\tablenotetext{4}{Mid-IR data from \citet{2009ApJ...699.1578M}}

\end{deluxetable}

\clearpage
\begin{deluxetable}{llrllcccc}
\tabletypesize{\scriptsize}
\tablecaption{ \label{table1} Lens Models}
\tablewidth{0pt}
\tablehead{
\colhead{Object}& \colhead{Ratio}& \colhead{Value}&\colhead{$R_{inner}$(kpc)\tablenotemark{*}}&\colhead{$R_{outer}$(kpc)\tablenotemark{*}}&\colhead{$\kappa_1$}&\colhead{$\kappa_2$}&\colhead{$\gamma_1$}& \colhead{$\gamma_2$}}
\startdata
HE 0047-1756&$\Delta m_{B-A}$&-0.19&$r_B=3.6$&$r_A=4.9$&$\kappa_A=0.43$&$\kappa_B=0.61$&$\gamma_A=0.48$&$\gamma_B=0.65$\\
HE 0435-1223&$\Delta m_{B-A}$&-0.24&$r_B=7.6$&$r_A=8.2$&$\kappa_A=0.46$&$\kappa_B=0.52$&$\gamma_A=0.39$&$\gamma_B=0.59$\\
&$\Delta m_{C-A}$&-0.30&$r_C=8.2$&$r_A=8.2$&$\kappa_A=0.46$&$\kappa_C=0.46$&$\gamma_A=0.39$&$\gamma_C=0.39$\\
&$\Delta m_{D-A}$&0.09&$r_D=7.0$&$r_A=8.2$&$\kappa_A=0.46$&$\kappa_D=0.56$&$\gamma_A=0.39$&$\gamma_D=0.64$\\
HE 0512-3329&$\Delta m_{B-A}$&-0.40&$r_A=2.2$&$r_B=3.3$&$\kappa_A=0.59$&$\kappa_B=0.41$&$\gamma_A=0.55$&$\gamma_B=0.37$\\
SDSS 0806+2006&$\Delta m_{B-A}$&-0.47&$r_A=7.2$&$r_B=3.0$&$\kappa_A=0.34$&$\kappa_B=0.82$&$\gamma_A=0.29$&$\gamma_B=0.77$\\
SBS 0909+532&$\Delta m_{B-A}$&-0.60&$r_A=3.4$&$r_B=6.3$&$\kappa_A=1.00$&$\kappa_B=0.55$&$\gamma_A=1.39$&$\gamma_B=1.19$\\
SDSS J0924+0219&$\Delta m_{B-A}$&0.00&$r_A=5.2$&$r_B=5.8$&$\kappa_A=0.50$&$\kappa_B=0.45$&$\gamma_A=0.45$&$\gamma_B=0.39$\\
FBQ 0951+2635&$\Delta m_{B-A}$&-0.69&$r_A=3.6$&$r_B=0.9$&$\kappa_A=0.28$&$\kappa_B=1.07$&$\gamma_A=0.15$&$\gamma_B=1.02$\\
QSO 0957+561&$\Delta m_{B-A}$&-0.30&$r_B=5.7$&$r_A=28.6$&$\kappa_A=0.20$&$\kappa_B=1.03$&$\gamma_A=0.15$&$\gamma_B=0.91$\\
SDSS J1001+5027&$\Delta m_{B-A}$&0.23&$r_A=10.6$&$r_B=5.0$&$\kappa_A=0.35$&$\kappa_B=0.74$&$\gamma_A=0.28$&$\gamma_B=0.72$\\
SDSS J1004+4112&$\Delta m_{B-A}$&0.00&---&---&$\kappa_A=0.48$&$\kappa_B=0.48$&$\gamma_A=0.59$&$\gamma_B=0.48$\\
&$\Delta m_{C-A}$&0.45&---&---&$\kappa_A=0.48$&$\kappa_C=0.38$&$\gamma_A=0.59$&$\gamma_C=0.33$\\
QSO 1017-207&$\Delta m_{B-A}$&-0.26&$r_A=5.4$&$r_B=1.5$&$\kappa_A=0.35$&$\kappa_B=1.23$&$\gamma_A=0.45$&$\gamma_B=1.32$\\
HE 1104-1805&$\Delta m_{B-A}$&0.60&$r_A=8.6$&$r_B=16.6$&$\kappa_A=0.64$&$\kappa_B=0.33$&$\gamma_A=0.52$&$\gamma_B=0.21$\\
PG 1115+080&$\Delta m_{A2-A1}$&-0.65&$r_{A1}=5.8$&$r_{A2}=5.9$&$\kappa_{A1}=0.49$&$\kappa_{A2}=0.51$&$\gamma_{A1}=0.44$&$\gamma_{A2}=0.55$\\
RXS J1131-1231&$\Delta m_{A-B}$&1.39&$r_A=10.1$&$r_B=10.2$&$\kappa_A=0.44$&$\kappa_B=0.43$&$\gamma_A=0.59$&$\gamma_B=0.51$\\
&$\Delta m_{C-B}$&1.58&$r_A=10.1$&$r_C=10.7$&$\kappa_B=0.43$&$\kappa_C=0.42$&$\gamma_B=0.51$&$\gamma_C=0.50$\\
SDSS J1206+4332&$\Delta m_{A-B}$&-0.56&$r_B=10.02$&$r_A=15.0$&$\kappa_A=0.43$&$\kappa_B=0.63$&$\gamma_A=0.41$&$\gamma_B=0.72$\\
SDSS J1353+1138&$\Delta m_{A-B}$&0.00&$r_B=1.6$&$r_A=5.2$&$\kappa_A=0.30$&$\kappa_B=0.96$&$\gamma_A=0.22$&$\gamma_B=0.89$\\
HE 1413+117\tablenotemark{**} &$\Delta m_{B-A}$&0.00&---&---&$\kappa_A=0.53$&$\kappa_B=0.43$&$\gamma_A=0.64$&$\gamma_B=0.34$\\
&$\Delta m_{C-A}$&-0.25&---&---&$\kappa_A=0.53$&$\kappa_C=0.46$&$\gamma_A=0.64$&$\gamma_C=0.35$\\
&$\Delta m_{D-A}$&-0.75&---&---&$\kappa_A=0.53$&$\kappa_D=0.58$&$\gamma_A=0.64$&$\gamma_D=0.69$\\
B J1422+231&$\Delta m_{A-B}$&0.16&$r_B=5.2$&$r_A=5.3$&$\kappa_A=0.38$&$\kappa_B=0.39$&$\gamma_A=0.53$&$\gamma_B=0.66$\\
&$\Delta m_{C-B}$&0.02&$r_B=5.2$&$r_C=5.7$&$\kappa_B=0.39$&$\kappa_C=0.36$&$\gamma_B=0.66$&$\gamma_C=0.48$\\
&$\Delta m_{D-B}$&-0.08&$r_B=5.2$&$r_D=1.3$&$\kappa_D=1.54$&$\kappa_B=0.39$&$\gamma_D=1.81$&$\gamma_B=0.66$\\
SBS 1520+530&$\Delta m_{B-A}$&-0.39&$r_A=9.5$&$r_B=3.0$&$\kappa_A=0.29$&$\kappa_B=0.90$&$\gamma_A=0.15$&$\gamma_B=0.85$\\
WFI J2033-4723&$\Delta m_{B-C}$&-0.50&$r_C=6.8$&$r_B=11.2$&$\kappa_B=0.38$&$\kappa_C=0.61$&$\gamma_B=0.25$&$\gamma_C=0.73$\\
&$\Delta m_{A2-A1}$&0.00&$r_{A2}=8.2$&$r_{A1}=9.3$&$\kappa_{A1}=0.48$&$\kappa_{A2}=0.55$&$\gamma_{A1}=0.39$&$\gamma_{A2}=0.65$\\
\enddata

\tablenotetext{*}{\, Computed using a concordance cosmology}
\tablenotetext{**}{\, Lens redshift unknown}

\end{deluxetable}

\clearpage

\begin{deluxetable}{lcccccccccc}
\tabletypesize{\scriptsize}
\tablecaption{ \label{table2} Frequency Distribution of Image Pairs, $n_{k_1,k_2}$}
\tablewidth{0pt}
\tablehead{
\colhead{}& \colhead{$k_2=0.35$}& \colhead{$k_2=0.45$}&\colhead{$k_2=0.55$}&\colhead{$k_2=0.75$}&\colhead{$k_2=0.85$}&\colhead{$k_2=1.05$}&\colhead{$k_2=1.15$}&\colhead{$k_2=1.25$}&\colhead{$k_2=1.55$}&\colhead{$k_2=1.75$}}
\startdata
$k_1=0.35$&$2$&1&---&$3$&$2$&$3$&---&$1$&$1$&---\\
$k_1=0.45$&---&$4$&$9$&1&---&---&---&---&---&---\\
$k_1=0.55$&---&---&$1$&---&---&1&---&---&---&---\\
\enddata

\end{deluxetable}

\end{document}